\documentclass[a4paper,10pt]{article}
\begin{document}

\title{Dream of a Christmas lecture}
\author{Alejandro Rivero\thanks{ 
Email: \tt rivero@wigner.unizar.es}}
\maketitle
\begin{abstract}
We recall the origins of differential calculus from a modern perspective. 
This lecture should be a victory song, but the pain makes it to sound more as a
oath for vendetta, coming from Syracuse two milenia before.
\end{abstract}

A visitor in England, if he is bored enough, could notice that our
old 20 pound notes are
decorated with a portrait of Faraday  
imparting the first series of "Christmas 
lectures for young people", which
began time ago, back in the XIXth century, at his suggestion. Today they
have become
traditional activity in the Royal Institution. 

This year the generic theme of the lectures was quantum theory and the 
limits implied by it. The BBC uses to broadcast the full sessions during
the holidays, and I decided to enjoy an evening seeing the recording.
This day, the third of the series, is dedicated to the time scale
of quantum phenomena. The main hall is to be occupied,
of course, by the children who have come to enjoy the experimental
session, and the BBC director, a senior well trained to control
this audience, keeps the attention explaining how the volunteers
are expected to enter and exit the scene. While he proceeds
to the customary notice, that "{\it all the demonstrations here are done
under controlled conditions and you should not try to repeat
them at home}", I dream of a zoom over a first bowl with some
of the bank notes, and the teacher starting the lecture.

He wears the white coat and in a rapid gesture drops a match in
the bowl, and the pieces of money take fire. The camera goes
from the flames to the speaker, who starts:

{\it Money. Man made, artificial, unnatural. Real and Untrue.} 

And then a slide of a stock market chart:

{\it But take a look
to this graph: Why does it move with the same equations that a
grain of pollen? Why does it oscillate as randomly as a quantum
mechanic system?}.

Indeed. It is already a popular topic that the equations used for the
derivative market are related to the heat equation, and there is
some research running in this address. But the point resonating in
my head was a protest, formulated\cite{shaw} a couple of months before by
Mr. W.T. Shaw, a researcher of financial agency, Nomura: 

"Money analysts get volatility and other parameters from the measured
market data, and this is done by using the inverse function theorem.
If a function has a derivative non zero in a point, then it is
invertible at this point. But, if we are working out discrete
calculus, if we are getting discrete data from the market, how
can we claim that the derivative is non zero? Should we say that
our derivative is {\it almost non zero}? What control do we have over
the inversion process?". 

Most meditations in this sense drive oneself to understand the hidings under
the concept of stability of a numerical integration process. 

But consider just this: discrete, almost zero, almost nonzero calculus!.
 It is a romantic concept by
itself. Infinitesimals were at the core of the greatest priority dispute
in Mathematics. On one side, at Cambridge, the second Lucasian chair,
Newton. On the other side, at the political service of the elector of
Mainz, the mathematical philosophy of Leibniz. And coming from 
the dark antiquity, old problems: {\it How do you get a straight line from
a circle?  How do you understand the area of any figure? What
is speed? Is the mathematical continuum composed of indivisible
"individua", mathematical "atoms without extension"?} 

Really all the thinking of calculus is pushed by two paradoxes. That one
of the volume and that one of the speed. 

The first one comes, it is said, from Democritus. Cut a cone with a plane
parallel and indefinitely  near to the basis. Is the circle on the plane
smaller or equal than the basis? 

Other version makes the infinite more explicit. Simply cut the cone parallel
to the basis. The circle in the smaller cone should be equal to the one
in the top of the trunk. But this happens for every cut, lets say you make
infinite cuts, always the circles will be equal. How is that different of
a cylinder? You can say, well, that the shape, the area, decreases between the
cuts, no in the cuts. Ok, good point. But take a slice bounded by two
cuts. As we keep cutting we make the slice smaller, indefinitely thinner,
until the distinction between to remove a slice and to make a cut is  
impossible. How can this distinction be kept? And we need to kept it in
a mathematically rigorous way, if possible.

The second paradox is a more popular one, coming from
the meditations of Zeno. In more than one sense, it is dual to the
previous one. Take time instead of height and position instead of circular
area. How can an arrow to have a speed? How can an arrow to change position
if it is resting at  every instant? In other version, it is say that it
can not move where it is fixed, and it can not move where it is not yet. Or,
as Garcia-Calvo, a linguist and translator of Greek, formulated once: 
"One does not kiss while he lives, one does not live when
he kisses".

Seriously taken, the paradox throws strong doubts about the concept of
instantaneous speed. Or perhaps about the 
whole conception of what "instantaneous" is. While Democritus asked how
indefinite parts of space could add up to a volume, Zeno wonders how a
movement can be decomposed to run across indefinite parts of time. A Wicked
interplay.  

It is interesting to notice that physicists modernly do not like to
speak of classical mechanics as a limit $\hbar \to 0$, but as a cancellation
of the trajectories that differ from the classical one. Perhaps this
is more acceptable. Anyway, 
the paradoxes were closed in false by Aristotle with some deep thoughts 
about the infinity.  Old mathematics was recasted for practical
uses and, at the end, lost. 

But in the late mid ages, some manuscripts were translated again. A man
no far from my homeland, in the Ebro river in Spain, took
over a Arabic book to be versed into Latin. It was the {\it Elements}, that 
book all you still "suffer" in the first courses of math in the primary school,
do you remember? Circles, angles, triangles, and all that. And, if your teacher
is good enough, the art of mathematical skepticism and proof comes
with it. 

Of course the main interesting thing in the mid ages is a new art, Algebra. 
But that is a even longer history. To us, our interest is that with
the comeback of geometry, old questions were again to be formulated. 
{\it If continuous becomes, in the limit, without extension, then is such
limit divisible? And if it is indivisible, atomic, how can it be?} 

That automatically brings up other deterred theory to compare with. That
one which postulates Nature as composed with indivisible atoms but, 
having somehow extension, or at least some vacuum between them.

Such speculation had begun to be resuscitated in the start of the XVII,
with Galileo Galilei himself using atomic theory to justify heat, colours,
smell. His disciple Vincenzo Viviani will write, time-after, that then, with
the polemic of the book titled "Saggiatore", the eternal prosecution 
of Galileo actions and discourses began.

Mathematics was needing also such atomic objects, and in fact the first
infinitesimal elements were named just that, atoms, before the modern
name was accepted. 

(By the way, Copernico in ``De revolutionibus'' explains how the atomic
model, with its different scales of magnitude, inspires the
astronomical world: the distance of the earth to the center
of the stars sphere is said to be negligible by inspiration
from the negligibility of atomic scale. It is very funny that some centuries 
later someone proposed the "planetary" model of atoms.)

Back to the lecture. Or to the dream. 
Now the laboratory  has activated a sort of TV projector
bringing images from the past. Italy.  

{\it Viviani. He made a good effort to recover Archimedes and other classical
geometers. So it is not strange that the would-to-be first lucasian,
Isaac Barrow, become involved when coming to Florence. And Barrow
understood how differentiation and integration are
dual operations.}

Noises...

{\it Perhaps Barrow learn of it during his Mediterranean voyage}

Noises
of swords and pirates sound here in the TV scene, and Barrow himself
enters in the lecture room. 

He is still blooding from the encounter with the pirates. Greets the
speaker, cleans himself, and smiles to
the children in the
first row: 

{\it "We become involved in a stupid war. Europe went to war about
sacraments, you know, the mystery of eucharistic miracle and all that 
niceties. And there we were, with individia, indivisilia, atoms... things
that  rule out difference between substance and accidents. You can not
make a bread into a divine body if it is only atoms, they say."} 

Indeed, someone filed a denounce against Galileo claiming that is
theory was against the dogma of transustantiation\cite{redondi}. Touchy 
matter, good
for protestant faith but not for the dogmas from Trent concilium. 

For a moment he raises the head, staring to us, in the upper circle. Then
he goes back to the young public: {\it "Yes, there was war.
 Protestants, Catholics, Anglicans.
Dogmas and soldiers across Europe. Bad time to reject Aristotle, worse
even to bring again Democritus. With Democritus comes Lucretius, with
Lucretius comes Epicurus. Politically inconvenient, you know. Do  the
answers pay the risk?"}

He goes away. He went away to Constantinople, perhaps to read the only
extant copy of the Archimedean law. {\it Perhaps he found the lost Method. 
Perhaps he lost other books when his ship was burned in Venice. }

Yes, Bourbaki says (according \cite{arnold}) that Barrow was the first one
proofing the duality between derivatives and integration. At least, with
his discrete "almost zero" differential triangle, doubting about the
 risks of jumping to
the limit, was closer to our modern \cite{connes} view. Three or four
years ago Majid, then
still in Cambridge, claimed its resurrection in the non commutative calculus 
$f(x) dx = dx f(x-\lambda)$. Even the formulation of fermions in the lattice,
according Luscher, depends on this relationship to proof the
cancellation of anomalies.

Also we would note that his calculus was "renormalized" to a finite
scale, as instead of considering directly $\Delta f/\Delta x$, he first
scaled this relationship to a finite triangle with side $f(x)$. The
freedom to choose either the triangle on $f(x)$ or the one in $f(x+\lambda)$
was lost when people start to neglect this finite scale.

Really, this is mathematical orthodoxy. Consider a series $\sin (1/n)$ and another
one $1/n$, both going to zero. The quotient, then, seems to go to an indefinite
$0/0$, but if you scale all the series to a common denominator, call it
$S_a(n)/a$, you will find that $S_a(n)$ goes to $a$ as $n$ increases. Wilson
in the seventies made the same trick for statistical field theory (or for
quantum field theory), which was at that moment crowded of problematic
infinities.

There is also a infinite there in the Barrow idea, but it is a very trivial
one. Just the relation between the vertical of the finite triangle and
the horizontal of the small one, $f(x)\over \Delta x$. It goes to infinity, but
this divergence can be cured by subtracting another infinite quantity,
$f(x+\lambda)\over \Delta x$, so that the limit is finite\footnote{This
example was provided by Alain in Vietri at the request of the public, but
it was not to be related to the hoft algebra of trees, as far as I can see}.

Barrow died in sanctity. But in his library \cite{feingold} there was no less
that three copies of Lucretius "De Rerum Natura", a romam poem about
atomistic Nature, already critiquized in the antiquity because in
supports the Epicurean doctrine: that gods, if they exist, are not
worried about the human affairs, so we must build our moral values
from ourselves and our relationships with our friends and society. 

In the lecture room, the slides fly one ager other. Back in the XVII,
with heat, smell, colour, and other accidents, 
black storms blow in the air. It has been proposed that the sacred
eucaristic mystery was in agreement with Aristotle, as it could be said
that the substance of wine and bread was substituted by the substance
of the Christ, while the accidents remained. Go tell to the Luterans.  

{\it First August 1632. The Compa\~nia de Jesus forbids the teaching of the
atomic doctrine. 22nd June 1633, Galileo recants}. ``Of all the days that 
was the one / An age of reason could have begun'' \cite{brecht}

In the "Saggiatore", Galileo
had begun to think of physical movement of atoms as the origin of the
heat. It would take three centuries for Einstein to get the Brownian
key. But even that was already disentangled of pure mathematics, so
it took some other half century for to discover the same equations
again, now for the stock market products. The history has not finished.

It sounds not to surprise that Dimakis has related discrete calculus
to the Ito calculus, the basic stochastic in the heat equation, the
play of money, that Black and Scholes rediscovered. In some sense
it is as if the physical world described by mathematics were dependent
on mathematics only, as it it were the unique answer to organise
things in a localized position. 

Dark clouds will block our view. {\it Barrow survived to his ship and
crossed Germany and come home to teach Newton.
But Newton himself missed
something greater when, for sake
of simplicity, the limit to zero was taken. In this limit, he can
claim the validity of series expansion to solve any differential
equation, so it is a very reasonable assumption. } Yes, but it had been
more interesting to control the series expansion even without such
limit.

Leibniz come to the same methods and the jump to the limit is
to be the standard. Mathematical atoms, scales and
discrete calculus will hide its interplay with the infinitesimal
ones for some centuries. Only two years ago Mainz, in voice of
Dirk Kreimer, got again the clue to generalized Taylor series. The
wood was found to be composed of trees.

Vietri is a small village in the Tyrrenian sea, near Salerno, looking at
the bay of Amalfi. Good fishing and intense {\it limoncello} liquor. About
the 20th of Mars, 1998, there Alain come, to explain the way Kreimer had 
found a Hopf algebra structure governing perturbative renormalization.
The algebra of trees was not only related to Connes Moscovici algebra,
but also with the old one proposed by Cayley to control the Taylor
series of the vector field differential equation.

And to close the circle, Runge-Kutta numerical integration algorithms 
can be
classified with a hoft algebra of trees. {\it Today it can 
be said\cite{lessons} that
the generic solution to a differential equation is not just the
function, but also some information codified in the Butcher
group}. Which can be related to the physics monster of this century, 
the renormalization group we have mentioned before. 

Can we control the inversion of the Taylor series using trees? Then
these doubts about the inverse function theorem in stock markets
could be sorted out. Will us be able to expand in more than
one variable? Still ignorabimus.

Worse, it is progressively clear that this kind of pre-Newtonian
calculus are a natural receptacle for quantum mechanics. Even the
stock market Ito equations are sometimes honoured as "Feynman-Kac-Ito"
formula, so marking its link with the quantum world. The difference
comes from the format of the time variable in both worlds. One should
think that time is more subtle than the intuitive "dot" that 
Newton put in the fluxion equations.

Perhaps we are now, then, simply correcting a flaw made three hundred
years ago. A flaw that Nature pointed to us, when it was clear the
failure of classical mechanics in the short scale.

But how did come to exist the conditions to such failure? Why did
geometry need to be reborn in the XVIIth century? Why did
the mathematicians so little information, so that the mistake has
a high probability\footnote{And, by the way, Why there was not the
slightest notion of probability in the old mathematics texts, so
they were unable even to consider it?} to happen?

If calculus, or "indivisilia", were linked to atomism already in the
old age, it could be a sort of explanation. Archimedes explain that
Democritus was the first finding "without mathematical proof" the volume
of the cone. And with Democritean science there was political problems
already in Roman times:

A leftist scholar, Farrington\cite{farrington}, claims
 that political stability was thought
to reside in some platonic tricks, lies, proposed
in "Republic": a solid set of unskeptic faith going up
the pyramid until the divine celestial gods.  Epicurus is seen
as a fighter for freedom, putting at risk social stability. Against
Plato ideals, Epicurus casts in his help the Ionian learning,
including Democritean mathematics and physics. According Farrington,
if government aspires to platonic republic, it must control or
suppress such kind of mathematics and physics. 

No surprise, if this is true, that the man who understood the floating
bodies and the centers of gravity, who stated the foundations
and integration, the process of mechanical discovery and mathematical
rigor, who was fervently translated by Viviani and Barrow, was killed
and dismissed. To be buried without a name, it could have been Archimedes'
own wish. But to get his books left out of the copy process for centuries
until there were extinct, that is a different thing.

{\it "Only a Greek copy of the Floating Bodies extant, found at Constantinople.
See here the palimpsest, the math almost cleared, a orthodox liturgy,
perhaps St John Chrisostom, wrote above instead."}

{\it "Let me to pass
the pages, and here you have, the only known version of Archimedes
letter about the Mechanical Method. Read only by three persons, perhaps
four, since it was deleted in the Xth century. Was this reading the
goal of Barrow in orient?"}

And even then, is it the same? It has been altered, the last occasion
in this century, when someone painted four evangelists over the Method. 

Hmm. Last Connes report \cite{rh1} quotes the Floating Bodies
principle, doesn't it?. More and more associations. Stop!

And recall.

Had our research been different if we had been fully aware of the
indivisilia problems, if we had tried hard for rigour? Perhaps. Only
in the XVI, rescued Apollonius and Archimedes, the new mathematics
re-taken the old issue. And, as we have seen, in a dark atmosphere.
Enough to confuse them and go into classical mechanics instead
of deformed mechanics. Instead of quantum mechanics.

The matter of copernicanism has been usually presented as a political
issue. Brecht made a brilliant sketch of it while staying in Copenhagen
with some friends, physicists which become themselves caught in the
dark side of our own century. We suspect that the matter of atomism
also has suffered because of this, and now it appears that
Differential Geometry
itself has run across a world of troubles since the assesination
of his founder in Sicily two milenia ago. The truth has been blocked
again and again by the status quo, by the "real world" preferring 
tales of stable knowledge to inquisitive minds learning
to crawl across, and with, the doubt.   

%I first heard of this matter 1991 in Leipzig, but ... winter karpatz
%...
%There I got the idea of fermions relating differentials, discussing with
%Julio Guerrero in the hotel room looking at the sea. The 31th we made
%the manif for the science in Madrid, the 1st I failed an interview
%with a local bank, the night from the 4th to the 5th I got my first kiss 

{\it If the goal of the Christmas lectures is to move young people to 
start a career in science, here is our statement: it is 
for the honour of human spirit. It is because understanding, reading the book
of nature, we calm our mind.
 Call it ataraxia, athambia, or simply tranquility.}

But we have been mistaken, wronged, delayed. The world has 
tricked, outraged, raped us. {\it When we have been wronged, should we not
to revenge?} Then our main  
motivation is here: when reality is a lie, the song of science
must be a song of vengeance. A man in Syracuse has been killed, 
all our milenia-old 
family has been dishonoured. Every mother, every child, every man in Sicily  
knows then the word. Vendetta. 

{\it Go to your blackboards, my children, and sing the song.
Just to clear any trace
of pain in the soul}.

%Escucha mi canto de victoria, de AGC

\end{document}